# The average submillimetre properties of Lyman-$\alpha$ Blobs at $z = 3$

N. K. Hine⋆,[1] J. E. Geach,[1] Y. Matsuda,[2,3] B. D. Lehmer,[4] M. J. Michałowski,[5] D. Farrah,[6]
M. Spaans,[7] S. J. Oliver,[8] D. J. B. Smith,[1] S. C. Chapman,[9] T. Jenness,[10]
D. M. Alexander[11], I. Robson[12], and P. van der Werf[13]

[1] *Centre for Astrophysics Research, School of Physics, Astronomy and Mathematics, University of Hertfordshire, College Lane, Hatfield, AL10 9AB, UK*
[2] *National Astronomical Observatory of Japan, 2-21-1 Osawa, Mitaka, Tokyo 181-8588, Japan*
[3] *The Graduate University for Advanced Studies (SOKENDAI), 2-21-1 Osawa, Mitaka, Tokyo 181-0015, Japan*
[4] *University of Arkansas, 226 Physics Building, Fayetteville, AR 72701, USA*
[5] *Institute for Astronomy, University of Edinburgh, Royal Observatory, Blackford Hill, Edinburgh, EH9 3HJ, UK*
[6] *Department of Physics, Virginia Tech, Blacksburg, VA 24061, USA*
[7] *Kapteyn Astronomical Institute, University of Groningen, P.O. Box 800, 9700 AV Groningen, Netherlands*
[8] *Astronomy Centre, Department of Physics and Astronomy, University of Sussex, Brighton BN1 9QH*
[9] *Department of Physics and Atmospheric Science, Dalhousie University, Halifax, NS, B3H 3J5, Canada*
[10] *Large Synoptic Survey Telescope Project Office, 933 N. Cherry Ave, Tucson, AZ, 85719*
[11] *Centre for Extragalactic Astronomy, Department of Physics, Durham University, South Road, Durham, DH1 3LE, UK*
[12] *UK Astronomy Technology Centre, Blackford Hill, Edinburgh EH9 3HJ*
[13] *Leiden Observatory, PO Box 9513, 2300RA Leiden, the Netherlands*

**ABSTRACT**

Ly$\alpha$ blobs (LABs) offer insight into the complex interface between galaxies and their circumgalactic medium. Whilst some LABs have been found to contain luminous star-forming galaxies and active galactic nuclei that could potentially power the Ly$\alpha$ emission, others appear not to be associated with obvious luminous galaxy counterparts. It has been speculated that LABs may be powered by cold gas streaming on to a central galaxy, providing an opportunity to directly observe the 'cold accretion' mode of galaxy growth. Star-forming galaxies in LABs could be dust obscured and therefore detectable only at longer wavelengths. We stack deep SCUBA-2 observations of the SSA22 field to determine the average 850$\mu$m flux density of 34 LABs. We measure $S_{850} = 0.6\pm0.2$mJy for all LABs, but stacking the LABs by size indicates that only the largest third (area $\geq 1794$ kpc$^2$) have a mean detection, at $4.5\sigma$, with $S_{850} = 1.4\pm0.3$mJy. Only two LABs (1 and 18) have individual SCUBA-2 $>3.5\sigma$ detections at a depth of 1.1mJy beam$^{-1}$. We consider two possible mechanisms for powering the LABs and find that central star formation is likely to dominate the emission of Ly$\alpha$, with cold accretion playing a secondary role.

**Key words:** galaxies: evolution - galaxies: protoclusters - galaxies: formation - galaxies: high-redshift.

## 1 INTRODUCTION

Lyman-$\alpha$ Blobs (LABs) are large diffuse regions of Lyman-$\alpha$ (Ly$\alpha$) emission (10 – 100 kpc scale) with integrated Ly$\alpha$ luminosities of $\sim 10^{42}$ - $10^{44}$ erg s$^{-1}$ typically found at $z = 2-6$ (although one was detected at $z \sim 1$, Barger et al. 2012). The first LAB was detected in the Small Selected Area 22$^{hr}$ (SSA22) field at $z \sim 3$ (Steidel et al. 2000), however, extended areas of diffuse Ly$\alpha$ emission (that we might now class as LABs) had earlier been detected around overdensities of luminous galaxies and AGN at $z \sim 2.4$ (Francis et al. 1996; Keel et al. 1999), in association with high redshift submillimetre galaxies (SMGs, Ivison et al. 1998) and around high redshift radio galaxies (De Breuck et al. 1999; Kurk et al. 2000). Later surveys have detected further high redshift LABs with a range of sizes and luminosities (Matsuda et al. 2004; Matsuda et al. 2009; Matsuda et al. 2011; Yang et al. 2009; Erb et al. 2011 Bridge et al. 2013). Many appear to lie in dense regions that are expected to become massive clusters (Steidel et al. 1998).

The Ly$\alpha$ emission in LABs is thought to be powered either by feedback processes (involving superwinds, massive stars or AGN), or cold gas accretion. Many LABs contain luminous ionizing sources that could provide the energy needed to generate the observed Ly$\alpha$ emission (Francis et al. 1996; Keel et al. 1999; Dey et al. 2005; Geach et al. 2005; Geach et al. 2007; Geach et al. 2009; Geach et al. 2014; Webb et al. 2009; Rauch et al. 2011; Cantalupo et al. 2012; Ao et al. 2015). One scenario is that Ly$\alpha$ is emitted from cold gas clouds in a central galaxy fuelled either by massive stars or an AGN. A fraction of the Ly$\alpha$ emission escapes the galaxy and is scattered into our line of sight by the circumgalactic medium (CGM, Zheng et al. 2010; Zheng et al. 2011; Steidel et al. 2011; Hayes et al. 2011b; Rauch et al. 2011; Cen & Zheng 2013; Geach et al. 2014). An alternative scenario involves ionizing radiation escaping from a galaxy, or AGN, (which may be offset from the centre of the LAB) leading to Ly$\alpha$ emission from cold gas in the CGM itself, which is then also scattered by gas in the CGM (Cantalupo et al. 2012; Prescott et al. 2015). Geach et al. (2009) showed that the typical bolometric luminosities of such sources are sufficient to power the observed Ly$\alpha$ emission. The third feedback process involves superwinds. Here a starburst leads to multiple supernovae, creating overlapping bubbles which form a superwind and shock heat gas (Taniguchi & Shioya 2000; Mori et al. 2004; Mori & Umemura 2006). It is possible that more than one of these processes is contributing to the observed emission, especially in the larger LABs that contain multiple galactic sources.

⋆ E-mail:n.hine@herts.ac.uk

However, not all LABs have been found to contain luminous galaxies or AGN (Nilsson et al. 2006; Smith & Jarvis 2007; Smith et al. 2008). Hydrodynamic simulations suggest that the growth of massive ($M_h \geq 12 M_\odot$) galaxies at $z \geq 2$ is dominated by 'cold mode' accretion. Narrow streams of cold ($T \sim 10^{4-5}$) pristine gas penetrate the hot, virially shocked gas in the galaxy halo (Katz et al. 2003; Keres et al. 2005; Keres et al. 2009; Dekel et al. 2009). So far there is little direct observational evidence for the existence of such flows, however simulations have predicted the emission of Ly$\alpha$ from infalling cold gas (Dijkstra et al. 2006b; Haiman et al. 2000). Goerdt et al. (2010) found that the cold streams in their high-resolution simulations could produce the observed physical properties of LABs and similar results were obtained by Rosdahl & Blaizot (2012). However the moving mesh code simulation investigated by Nelson et al. (2013) did not support the existence of cold flows in all dark matter haloes (DMHs). They found that the fraction of gas that remained cold as it approached the central galaxy was sensitive to the simulation code used. In addition Faucher-Giguère et al. (2010) found that in their simulations cold accretion could not power the LABs unless emission from dense cores capable of producing star formation were included, whereas feedback processes could provide the observed emission.

It is difficult to predict the actual Ly$\alpha$ flux resulting from the gas flows generated in these simulations. The flux will depend on gas turbulence, radiative transfer and the presence of local ionizing sources (Rosdahl & Blaizot 2012). The radiative transfer is particularly complicated due to the resonance of the Ly$\alpha$ line (Neufeld 1990) and its sensitivity to assumptions about sub-grid physics (Nelson et al. 2013). Simulations using different models of radiative transfer therefore predict a wide range of physical and observable properties (Dijkstra et al. 2006a; Dijkstra et al. 2006b; Faucher-Giguère et al. 2010; Rosdahl & Blaizot 2012; Nelson et al. 2013).

Recently Prescott et al. (2015) suggested that the non-detection of a luminous galaxy within some LABs is actually evidence *against* cold accretion, as such flows should be triggering star formation at a similar rate to the gas inflow ($\sim$100s $M_\odot$ yr$^{-1}$). Hayes et al. (2011a) found evidence of polarization from a large

LAB at $z \sim 3$, suggesting the presence of a central ionizing source rather than cold accretion (see also Geach et al. 2014). Cen & Zheng (2013) presented a model relying primarily on ionizing sources to provide the energy for LABs, but with a contribution from cold gas accretion. This successfully reproduced the Ly$\alpha$ luminosity function and the luminosity-size relation of the Matsuda et al. (2004) and Matsuda et al. (2011) LABs. It appears likely that both cold gas accretion and heating by feedback processes play a part in the creation of LABs, but the relative importance of the different mechanisms is still unclear.

In this work we consider the LABs in the extensively studied $z = 3.1$ SSA22 protocluster at RA = 22h 17m, Dec = +00$^\circ$ 15$'$ (Steidel et al. 1998; Steidel et al. 2000; Steidel et al. 2003; Hayashino et al. 2004; Matsuda et al. 2005; Lehmer et al. 2009; Weijmans et al. 2010; Yamada et al. 2012a; Yamada et al. 2012b). The two largest LABs in SSA22 (LAB1 and LAB2) were first detected by Steidel et al. (2000) during a general Ly-$\alpha$ survey of the protocluster. LAB1 and LAB2 have since been studied extensively at a range of wavelengths (Chapman et al. 2001; Chapman et al. 2004; Bower et al. 2004; Wilman et al. 2005; Geach et al. 2005; Geach et al. 2014; Matsuda et al. 2007; Webb et al. 2009; Uchimoto et al. 2008; Uchimoto et al. 2012; Martin et al. 2014). A comprehensive search for LABs in the protocluster (Matsuda et al. 2004, M04) identified a total of 35 LABs with a flux greater than $0.7 \times 10^{-16}$ erg s$^{-1}$ cm$^{-2}$ (which they defined as the limit for LABs) varying in area isophotal from 222 arcsec$^2$ to 16 arcsec$^2$.

Geach et al. (2005) used the Submillimetre Common User Bolometer Array (SCUBA) instrument on the James Clerk Maxwell Telescope (JCMT) to search for submillimetre (submm) sources in 25 of the LABs. They found individual 850$\mu$m detections at $\geq 3.5\sigma$ for five LABs (LABs 1,5,10,14,18, IDs from M04) and a detection at the $\geq 3.0\sigma$ level for the full sample using mean stacking (at a $1\sigma$ depth of 1.5mJy for the main sample and $\sim$5.3mJy for a subset). However, AzTEC/ASTE 1.1mm observations of the 35 LABs (with an rms noise level of $0.7 - 1$ mJy beam$^{-1}$) did not detect any sources at $>3.5\sigma$, whilst their stacking analysis showed a mean $S_{1.1mm}$ <0.40 mJy ($3\sigma$) (Tamura et al. 2013). More recently, deeper Atacama Large Millimeter/submillimeter Array (ALMA) observations have detected 1.1mm sources in LABs 12 and 14 (Umehata et al. 2015).

The deeper SCUBA-2 survey of the SSA22 field (1.1 mJy, compared to 1.5 mJy at $1\sigma$) is now complete and we have used these data to revisit the work of Geach et al. (2005) obtaining updated individual and stacked submm detections for 34 SSA22 LABs (there is no coverage of LAB17). SCUBA-2 sources at $z \sim 3$ are expected to be SMGs which could provide the star-forming activity required to fuel the LABs.

Details of the SCUBA-2 and *Spitzer* observations used in our work are described in section 2. Our results are set out in section 3, compared to theoretical predictions in section 4 and discussed further in section 5. We assume a $\Lambda$ cold dark matter ($\Lambda$CDM) cosmology of $\Omega_m = 0.31$, $\Omega_\Lambda = 0.69$ and $H_0 = 68$ km s$^{-1}$ Mpc$^{-1}$ giving an angular scale at $z \sim 3.1$ of 7.8 kpc per arcsec.

## 2 OBSERVATIONS

### 2.1 Submm observations

The SSA22 field was observed as part of the JCMT Submillimetre Common User Bolometer Array 2 (SCUBA-2, Holland et al. 2013 Cosmology Legacy Survey (S2CLS, project ID MJLSC02). 105 observations were made between 23 August 2012 and 29 November 2013 to produce a 30' diameter map centred on 22:17:36.3, +00:19:22.7. The limiting conditions were a zenith optical depth in the range $0.05 < \tau_{225} < 0.1$, with a mean $\tau_{225} = 0.07$. The beam-convolved map has a $1\sigma$ depth of 1.1 mJy beam$^{-1}$ and an integration time of ~3000s per 2 arcsec pixel.

The data reduction steps are described fully in Geach et al. (2016 in preparation), but we describe the main steps here. The Dynamical Iterative Map-Maker (DIMM) within the *Sub-Millimetre Common User Reduction Facility* (SMURF; Chapin et al. 2013) is used to extract astronomical signals from each SCUBA-2 bolometer time stream, mapping the result onto a celestial projection. All S2CLS maps are projected on a tangential co-ordinate system with 2 arcsec pixels.

Flat-fields are applied to the time-streams using flat scans that bracket each observation, and a polynomial baseline fit is subtracted from each time stream. Data spikes are rejected (using a $5\sigma$ threshold in a box size of 50 samples), DC steps are removed and gaps filled. Next, an iterative process begins that aims to fit the data with a model comprising a common mode signal, astronomical signal and noise. The common mode modelling is performed independently for each SCUBA-2 sub-array, deriving a template for the average signal seen by all the bolometers; it is removed from the stream, and an extinction correction is applied (Dempsey et al. 2013). Next, a filtering step is performed in the Fourier domain, which rejects data at frequencies corresponding to angular scales $\theta > 150$ arcsec and $\theta < 4$ arcsec. Finally, a model of the astronomical signal is determined by gridding the time streams onto a celestial projection (since a given sky position will have been visited by many independent bolometers) and then subtracted from the input time streams. The iterative process continues until the residual between the model and the data converges.

The last processing step is to apply a matched filter to the maps, convolving with the instrumental PSF to optimize the detection of point sources. We use the PICARD recipe *scuba2_matched_filter* which first smooths the map (and the PSF) with a 30 arcsec gaussian kernel, then subtracts this from both to remove any large scale structure not eliminated in the filtering steps that occurred during the DIMM reduction. The map is then convolved with the smoothed beam. A flux conversion factor of 591 Jy beam$^{-1}$ pW$^{-1}$ is applied; this canonical calibration is the average value derived from observations of hundreds of standard submillimetre calibrators observed during the S2CLS campaign (Dempsey et al. 2013), and includes a 10 per cent correction necessary to account for losses that occur due to the combination of filtering steps we apply to the data (see Geach et al. 2013). The flux calibration is expected to be accurate to within 15 per cent.

### 2.2 Infrared observations

We obtained reduced *Spitzer Space Telescope* (*Spitzer*) observations of the SSA22 protocluster from the Infrared Science Archive (IRSA). The observations used were from GTO 64 and GO 30328 (P.I.: Fazio) and GO 3473 (P.I.: Blain). PID 64 and 30328 were used by Webb et al. (2009) in their analysis of IRAC and MIPS sources in the SSA22 LABs. PID 64 is a single pointing consisting of a ~ $5' \times 5'$ area observed with IRAC (AORID 4397824) with an integration time of 6400s pixel$^{-1}$. PID 30328 covered an area of ~ 375 arcmin$^2$ using IRAC (AORIDs 17599488, 17599744, 17600000, 17600256, 17600512) with integration times of $3000 - 7500$s pixel$^{-1}$. This included a 225 arcmin$^2$ region covered by all four wavelengths to a uniform depth of 7500s pixel$^{-1}$. We used data from PID 3473 for LABs not covered by these deep observations. These shallower observations achieved a depth of 0.2, 0.5, 3.1 and 4.5$\mu$Jy for IRAC channels 1 to 4 respectively (Hainline et al. 2009). Where available we reviewed data for all four IRAC channels (3.6 - 8 $\mu$m). These data were used to identify IRAC sources within the LABs, the coordinates of which were then used for stacking the SCUBA-2 data, see Section 3.2.

## 3 ANALYSIS

### 3.1 Individual Sources

The SCUBA-2 850 $\mu$m flux density ($S_{850}$) measurements for the individual LABs are listed in Table 1. Only two have significant detections at $\geq 3.5\,\sigma$, LAB1 and LAB18, indicated in bold in the Table. LAB16 is marginally detected at $\sigma = 3.0$.

Our results differ from those in Geach et al. (2005, G05), which had five detections at $\geq 3.5$ sigma and larger flux density values for both LAB1 and LAB18. The revised flux density figure for LAB1 was first reported in Geach et al. (2014), which suggested that the discrepancy may have been due to flux boosting in the original SCUBA data (Chapman et al. 2001). As well as flux boosting, the SCUBA results from G05 may be subject to enhancement due to early issues with data reduction and calibration, whereas the SCUBA-2 pipeline is considered to be more mature and more reliable. There is no SCUBA-2 coverage of LAB17 which lies on the edge of the field.

### 3.2 Stacking

We stacked the SCUBA-2 data to determine whether on average a submm source lies within each LAB. This would indicate a dust obscured source that could potentially produce the observed Ly$\alpha$ emission in LABs where no optical source has been detected. The stack consists of the inverse variance weighted mean of each pixel value in a $31 \times 31$ pixel square centred on the coordinates of each LAB (see Table 2). The value of the central pixel in the stack was taken as the weighted mean flux of all 34 LABs. The large size of the LABs (especially LAB1 and LAB2) introduces the risk of missing any associated SCUBA-2 sources, as the point of maximum $L_{Ly\alpha}$ emission (M04 coordinates) may not coincide with the SMG responsible for the emission due to scattering. This risk is mitigated to some extent by the large beam size (15 ''), but to reduce it further we used data from *Spitzer* to search for infrared sources lying within, or close to the LABs (Geach et al. 2007; Webb et al. 2009). IRAC sources are often used to identify SMG counterparts as these longer wavelengths are less effected by dust than optical or UV (Ashby et al. 2006; Pope et al. 2006; Biggs et al. 2011; Michałowski et al. 2012; Koprowski et al. 2015). By stacking on the coordinates of *Spitzer* sources (see Table 1), where available, we increased our chances of stacking on the location of any SMG within the LAB. We identified potential IRAC counterparts using an aperture of radius = area$^{1/2}$ (the isophotal area of each LAB is as given in Table 1) centred on the original M04 coordinates. Where there was more than one potential source (e.g LAB1) we used existing studies to confirm the appropriate coordinates where possible (Geach et al. 2007; Webb et al. 2009) We fitted a gaussian to the IRAC photometry to obtain the IRAC coordinates used in our stacking.

**Table 1.** Summary of Lyman alpha and 850 $\mu$m data relating to the individual SSA22 Lyman Alpha Blobs. The coordinates in columns two and three are from M04 and give the location of the maximumm Ly$\alpha$ luminosity. The IRAC coordinates give the offset from these values. The higher rms for LAB17 is because it lies on the edge of the map. The G05 results are included for comparison. [1] The submm source close to LAB8 falls within the original boundary of LAB1. [2] Data taken from M04.

| ID[2] | RA[2] | Dec[2] | Area[2] arcsec[2] | IRAC $\Delta\alpha$ '' | IRAC $\Delta\delta$ '' | Log(L$_{Ly\alpha}$)[2] erg s$^{-1}$ | S$_{850}$ mJy | G05 S$_{850}$ mJy |
|---|---|---|---|---|---|---|---|---|
| **LAB1** | **22 17 26** | **+00 12 32** | **222** | **3.3** | **3.0** | **44.0** | **4.6±1.1** | **16.8±2.9** |
| LAB2 | 22 17 39 | +00 13 23 | 152 | 3.7 | 7.9 | 43.9 | 0.1±1.1 | 3.3±1.2 |
| LAB3 | 22 17 59 | +00 15 25 | 78 | 1.4 | 4.3 | 43.8 | 0.1±1.1 | -0.2±1.5 |
| LAB4 | 22 17 25 | +00 22 05 | 57 | 4.3 | -1.2 | 43.6 | 2.4±1.1 | 0.9±1.5 |
| LAB5 | 22 17 12 | +00 16 41 | 55 | 2.0 | 4.7 | 43.2 | 1.9±1.1 | 5.2±1.4 |
| LAB6 | 22 16 51 | +00 24 58 | 42 | 1.4 | 5.7 | 43.2 | 1.0±1.1 | -0.5±1.8 |
| LAB7 | 22 17 41 | +00 11 20 | 40 | no source | no source | 43.2 | 1.2±1.1 | 0.3±1.6 |
| LAB8 | 22 17 26 | +00 12 50 | 39 | 2.2 | -2.7 | 43.2 | 2.6±1.1[1] | 0.3±5.3 |
| LAB9 | 22 17 41 | +00 17 20 | 38 | 6.0 | -1.5 | 43.1 | 2.2±1.1 | 1.3±5.3 |
| LAB10 | 22 18 02 | +00 25 52 | 34 | 1.3 | 7.4 | 43.3 | 3.2±1.1 | 6.1±1.4 |
| LAB11 | 22 17 20 | +00 17 28 | 30 | -1.5 | -1.2 | 43.0 | 2.5±1.1 | -0.4±5.3 |
| LAB12 | 22 17 32 | +00 16 55 | 29 | 3.8 | 1.3 | 42.9 | 0.8±1.1 | 3.2±1.6 |
| LAB13 | 22 18 08 | +00 16 41 | 28 | -1.8 | 5.6 | 43.0 | 1.2±1.1 | - |
| LAB14 | 22 17 36 | +00 15 54 | 27 | -0.8 | -0.5 | 43.1 | 2.0±1.1 | 4.9±1.3 |
| LAB15 | 22 18 08 | +00 10 19 | 26 | -2.5 | 1.8 | 43.3 | 2.4±1.1 | - |
| LAB16 | 22 17 25 | +00 11 13 | 25 | 2.6 | 5.3 | 43.0 | 3.4±1.1 | 2.2±5.3 |
| LAB17 | 22 18 36 | +00 07 16 | 24 | 0 | 0 | 43.1 | - | - |
| **LAB18** | **22 17 29** | **+00 07 48** | **22** | **3.1** | **-3.6** | **42.9** | **5.2±1.1** | **11.0±1.5** |
| LAB19 | 22 17 19 | +00 18 43 | 21 | 2.8 | 2.4 | 43.1 | -0.4±1.1 | -8.6±5.3 |
| LAB20 | 22 17 35 | +00 12 43 | 21 | no source | no source | 42.8 | 0.2±1.1 | 0.4±1.5 |
| LAB21 | 22 18 17 | +00 12 04 | 20 | 0.7 | 3.1 | 42.9 | 0.9±1.1 | - |
| LAB22 | 22 17 35 | +00 23 31 | 20 | no source | no source | 42.9 | 1.3±1.1 | - |
| LAB23 | 22 18 08 | +00 23 13 | 19 | no source | no source | 43.0 | 1.0±1.1 | - |
| LAB24 | 22 18 01 | +00 14 35 | 19 | no source | no source | 42.9 | -0.6±1.1 | - |
| LAB25 | 22 17 22 | +00 15 47 | 19 | no source | no source | 42.8 | -1.5±1.1 | 1.4±5.3 |
| LAB26 | 22 17 28 | +00 17 28 | 18 | -3.0 | 2.0 | 42.8 | 1.1±1.1 | -2.7±5.3 |
| LAB27 | 22 17 07 | +00 21 25 | 18 | -1.6 | 1.4 | 42.8 | 2.1±1.1 | 0.5±1.6 |
| LAB28 | 22 17 59 | +00 22 48 | 18 | no source | no source | 43.3 | -0.6±1.1 | - |
| LAB29 | 22 16 54 | +00 22 55 | 17 | no source | no source | 42.8 | 0.7±1.1 | - |
| LAB30 | 22 17 32 | +00 11 28 | 17 | -1.9 | 0.6 | 43.0 | 1.9±1.1 | 3.3±1.3 |
| LAB31 | 22 17 39 | +00 10 59 | 17 | no source | no source | 43.0 | 0.0±1.1 | -3.7±5.3 |
| LAB32 | 22 17 24 | +00 21 50 | 17 | no source | no source | 43.0 | 0.9±1.1 | 1.8±1.4 |
| LAB33 | 22 18 12 | +00 14 28 | 16 | no source | no source | 43.0 | 0.7±1.1 | 1.6±1.5 |
| LAB34 | 22 16 58 | +00 24 25 | 16 | no source | no source | 42.9 | 0.4±1.1 | - |
| LAB35 | 22 17 25 | +00 17 13 | 16 | -4.0 | 1.7 | 43.0 | 1.0±1.1 | 1.2±5.3 |

We also stacked subgroups of LABs to explore the possibility that they are not a homogeneous population, as more than one process may be responsible for the L$_{Ly\alpha}$ emission. We stacked based on both area and L$_{Ly\alpha}$. The results of our various stacks are presented in Table 2 (which also includes the corresponding mean L$_{Ly\alpha}$ and star formation rate for each stack) and Fig. 3.

Stacking all 34 LABs gives a significant detection of 0.6±0.2 mJy at 3.1$\sigma$, compared to a higher value of 3.0±0.9 mJy at 3.3$\sigma$ in G05. We also stacked just those LABs included in the G05 stack resulting in a 0.8±0.2mJy detection at 3.5$\sigma$. This suggests that the smaller sample in G05 may have introduced some bias, but most of the change is due to the improved data. We compare our work to that of G05 in Section 5.2. Excluding the two individually detected LABs reduces the stacked result to 0.3±0.2mJy at 1.7$\sigma$. The validity of excluding these two individual detections is controversial, but we include the result for information. We also carried out blind stacking of 10,000 sets of 34 random coordinates. This indicated a less than 0.1 per cent chance of obtaining a $\geq 3\sigma$ detection.

Fig. 1 presents a 62''× 62'' (∼480kpc projected at z ∼3.1)

image showing the result of stacking all 34 LABs. The contours start at 2$\sigma$ and are in steps of 0.5$\sigma$. Fig. 2 shows a histogram of the individual S$_{850}$ values and the associated uncertainties for each bin. This confirms that the average S$_{850}$ value for all LABs is greater than zero.

Stacking by size indicates a significant difference between the LABs with the largest areas and those with the smallest areas. LABs 1-12 (large LABs, > 29 arcsec$^2$) have a significant 4.5$\sigma$ detection (3.4$\sigma$ if we exclude LAB1), whilst LABs 13-24 (medium LABs, 28 - 19 arcsec$^2$) and LABs 25-35 (small LABs $\leq$ 19 arcsec$^2$) do not have a significant detection (<2$\sigma$ for both groups). The Ly$\alpha$ Bright (L$_{Ly\alpha}$ > 1.0 × 10$^{43}$ erg s$^{-1}$) stack has a marginal 3.0$\sigma$ detection (2.1$\sigma$ excluding LAB1) whilst the Ly$\alpha$ Faint (L$_{Ly\alpha}$ < 1.0 × 10$^{43}$ erg s$^{-1}$) LABs are not detected (1.6$\sigma$, falling to 0.8$\sigma$ when LAB18 is excluded). This indicates that the Ly$\alpha$ luminosity may be correlated with the S$_{850}$. However, neither population has a significant detection at $\geq$3.5$\sigma$ and the result is very dependent on whether LAB16 is included in the faint or bright group (LAB16 is the the 18th brightest of all the LABs, so is in-cluded in the bright group only because we don't have coverage for LAB17 and could otherwise be included in either group). These differences are also illustrated in Fig. 3 which shows the $S_{850}$ values for each stack. The large LABs have a marginally higher submm flux density than the other groups, albeit with a ∼1 $\sigma$ significance.

**Table 2.** LAB stacking results (LAB17 is not included in any of the stacks as there is no data for this LAB.) The mean $L_{Ly\alpha}$ for each stack is also included, based on the figures in M04. Star formation rates (SFR) are calculated from $S_{850}$ using a modified black body model (see Section 4.1).

| Stacked Sample | $S_{850}$ mJy | Mean $L_{Ly\alpha}$ $\times 10^{43}$ erg s$^{-1}$ | SFR M$_\odot$ year$^{-1}$ |
|---|---|---|---|
| All | 0.6±0.2 | 1.8±0.05 | 50±20 |
| All non detections (excluding LAB1 and LAB18) | 0.3±0.2 | 1.6±0.05 | 30±20 |
| LABs 1-12 (> 29 arcsec$^2$) | 1.4±0.3 | 3.1±0.09 | 120±30 |
| LABs 2-12 (152 - 29 arcsec$^2$) | 1.2±0.3 | 2.7±0.09 | 100±30 |
| LABs 13-24 (28 - 19 arcsec$^2$) | 0.6±0.3 | 1.0±0.09 | 50±30 |
| LABs 13-24 (excl. 18, 28 - 19 arcsec$^2$) | 0.2±0.4 | 1.0±0.1 | 20±70 |
| LABs 25-35 (19 - 16 arcsec$^2$) | -0.3±0.3 | 0.9±0.09 | - |
| Ly$\alpha$ Bright ($L_{Ly\alpha} > 1.0 \times 10^{43}$ erg s$^{-1}$) | 0.8±0.3 | 2.9±0.07 | 70±30 |
| Ly$\alpha$ Faint ($L_{Ly\alpha} < 1.0 \times 10^{43}$ erg s$^{-1}$) | 0.4±0.3 | 0.8±0.07 | 30±30 |

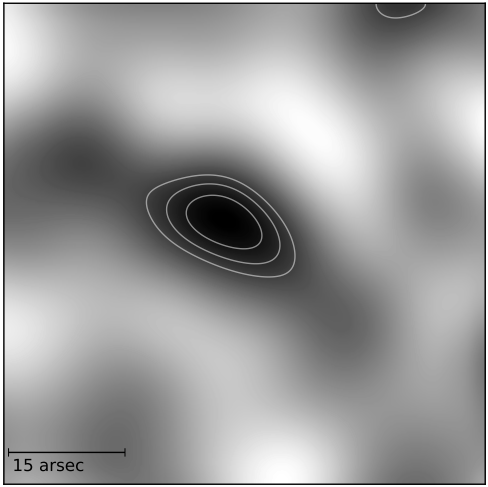

**Figure 1.** Result of stacking all 34 LABs. The image is $62'' \times 62''$ and the contours start at $2\sigma$ and are in steps of $0.5\sigma$ (dashed contours are negative $2\sigma$).

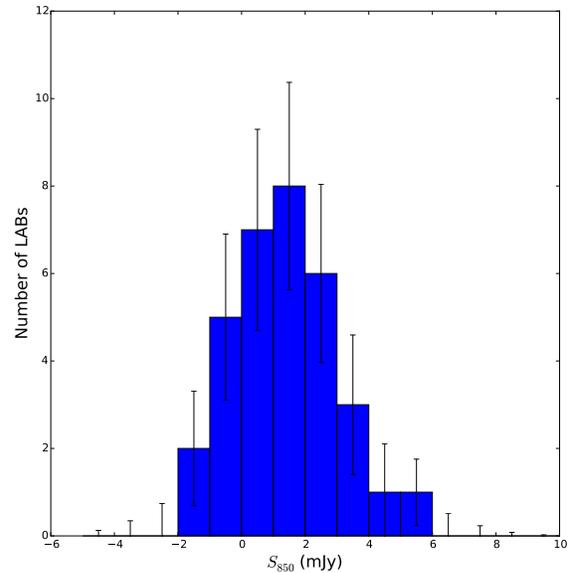

**Figure 2.** A histogram of the SCUBA-2 $S_{850}$ values for all LABs. The error bars were calculated using Monte Carlo Sampling.

## 4 COMPARISON TO THEORETICAL MODELS

We now compare our SCUBA-2 results to two possible models for the production of LABs, central star formation (where the Ly$\alpha$ emission is generated inside a central galaxy and then scattered in the CGM) and cold gas accretion.

### 4.1 Star formation model

We assumed that the far infrared SED can be modelled by a modified black body allowing us to calculate theoretical values for total infrared luminosity ($L_{IR}$) corresponding to given $S_{850}$ values. We used the modified blackbody function in equation 1 (Hilde-brand 1983; Magnelli et al. 2012) and integrated between 8$\mu$m and 1000$\mu$m (rest frame).

$$S_v \propto \frac{v^{3+\beta}}{\exp(hv/kT_{dust}) - 1} \qquad (1)$$

where $S_v$ is flux density, $\beta$ is the dust emissivity spectral index (assumed to be 2) and $T_{dust}$ is the dust temperature (assumed to be 30K). We discuss the appropriateness of our assumed values in Section 5.3.

A correlation between the star formation rate (SFR) and $L_{IR}$ is given in Kennicutt & Evans (2012) (see also Hao et al. 2011; Murphy et al. 2011).

$$\log(SFR_{IR}/M_\odot yr^{-1}) = \log(L_{IR}/erg s^{-1}) - 43.41 \qquad (2)$$

A similar relationship between the intrinsic Ly$\alpha$ star formation rate (SFR$_{Ly\alpha i}$) and intrinsic Ly$\alpha$ luminosity ($L_{Ly\alpha i}$) can be obtained (equation 3) by combining the standard Case B $L_{Ly\alpha}$: Hydrogen-$\alpha$ luminosity ($L_{H\alpha}$) ratio (8.7, Brocklehurst 1971) and

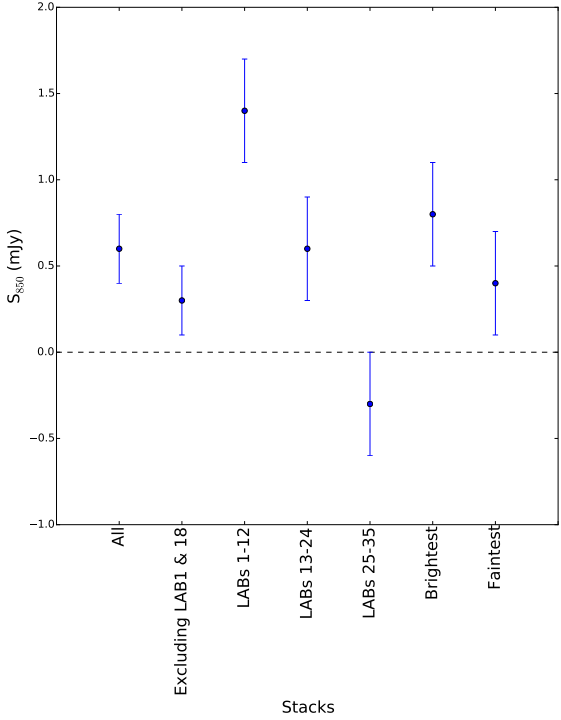

**Figure 3.** The SCUBA-2 $S_{850}$ values for the key stacks are presented. The first two points illustrate the full stack with and without the two individual detections. The next three points relate to the LABs' size and the final points to the LABs' brightness. This illustrates the higher flux densities associated with the larger LABs.

the Kennicutt & Evans (2012) $L_{H\alpha}$ to SFR correlation (equation 4, Dijkstra & Westra 2010):

$$\log(SFR_{Ly\alpha}) = \log(L_{Ly\alpha i}/8.7) - 41.27 \quad (3)$$

$$\log(SFR_{H\alpha}) = \log(L_{H\alpha}) - 41.27 \quad (4)$$

which implies

$$L_{Ly\alpha i} \approx 0.06 \times L_{IR} \quad (5)$$

This assumes $L_{IR}$ scales linearly with $S_{850}$, all $L_{IR}$ is from re-processed starlight (representing the majority of active star formation) and a Kroupa initial mass function (IMF, Kroupa & Weidner 2003).

To obtain the observed $L_{Ly\alpha}$ we must consider the fraction of $Ly\alpha$ photons that actually escape from the galaxy, $f_{esc}$, and the fraction of these photons that are then scattered by the CGM into our line of sight, $f_{sca}$ (Geach et al. 2014).

$$L_{Ly\alpha} \approx 0.06 \times f_{esc} \times f_{sca} \times L_{IR} \quad (6)$$

In Fig. 4 (upper image) we plot $L_{Ly\alpha}$ against $S_{850}$ for the individual LABs, the stacked LABs (All 34 LABs), the area stacks (large LABs, medium LABs and small LABs) and the two individually detected LABs (LAB1 and 18). We also indicate the theoretical position of LABs resulting from star formation processes, as calculated above, for a range of values of $f_{esc}f_{sca}$.

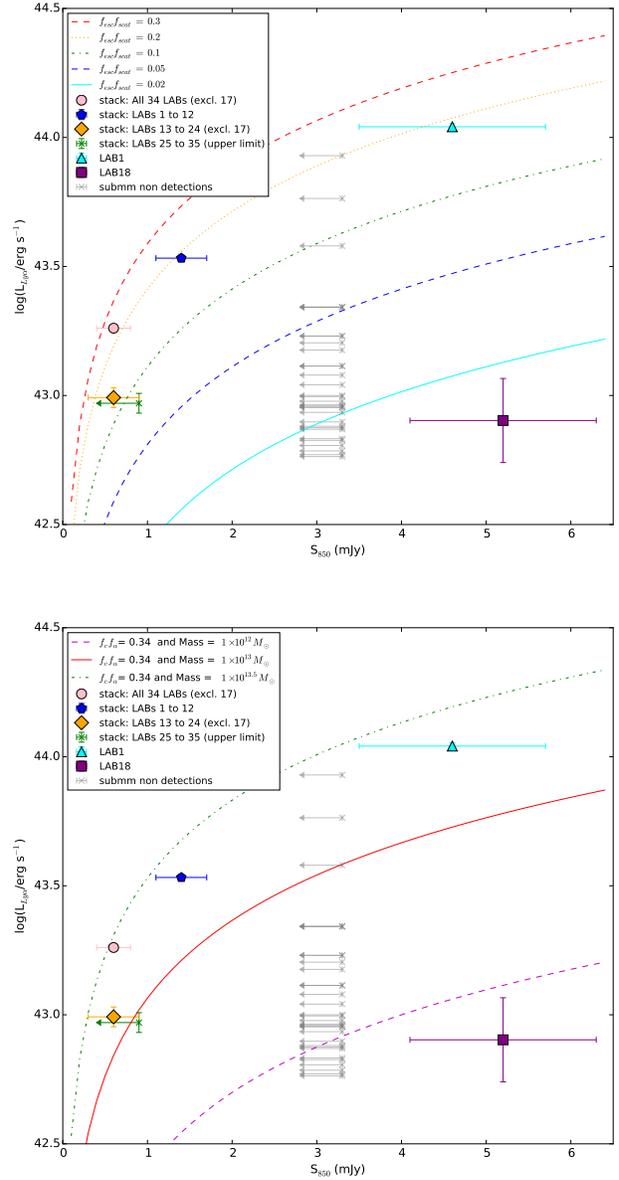

**Figure 4.** Log $Ly\alpha$ luminosity plotted against $S_{850}$ for the full stack (pink circle), large LABs (blue pentagon), medium LABs (orange diamond), small LABs (green arrow upper limit only) and the two individually detected LABs, LAB1 (cyan triangle) and LAB18 (purple square). Non detections are shown as grey upper limits at $3 \times$ rms. The lines in the upper image indicate the theoretical $L_{Ly\alpha}$-$S_{850}$ relationship if the LABs are fuelled by star formation processes for a range of values of $f_{esc}f_{sca}$. In the lower image the lines indicate the theoretical $L_{Ly\alpha}$-$S_{850}$ relationship if the LABs are fuelled by cold accretion, for a range of values of halo mass with $f_c f_\alpha = 0.34$.

Hayes et al. (2011c) combined the results of a number of observational studies out to $z \sim 8$ to produce an expression for the evolution of the $L_{Ly\alpha}$ escape fraction with redshift, giving a range of values, 5 – 10 per cent at $z \sim 3$. This is in reasonable agreement with the Dijkstra & Jeeson-Daniel (2013) empirical constraint of the effective $L_{Ly\alpha}$ escape fraction at $\sim 10$ per cent for $z \sim 3 - 4$. These papers define the effective $L_{Ly\alpha}$ escape fraction as the ratio of ob-served to intrinsic $L_{Ly\alpha}$ and hence their figures includes both our $f_{esc}$ and $f_{sca}$. We have plotted theoretical lines with $f_{esc}f_{sca} = 0.02$, 0.05, 0.1, 0.2 and 0.3. LAB18 can easily be fuelled with an effective escape fraction of 0.02, well below the values in the literature. This rather low value could be due to a more homogeneous gas distribution restricting the escape of Ly$\alpha$, or higher levels of dust obscuring the emission. LAB1, the full stack and large LABs require a higher value of $f_{esc}f_{sca}$ 0.2-0.3, whilst the small and medium LABs could be produced with a lower value, closer to that in the literature. If the escape fractions in the literature are correct, this suggests that central star formation alone may not be sufficient to fuel the larger LABs.

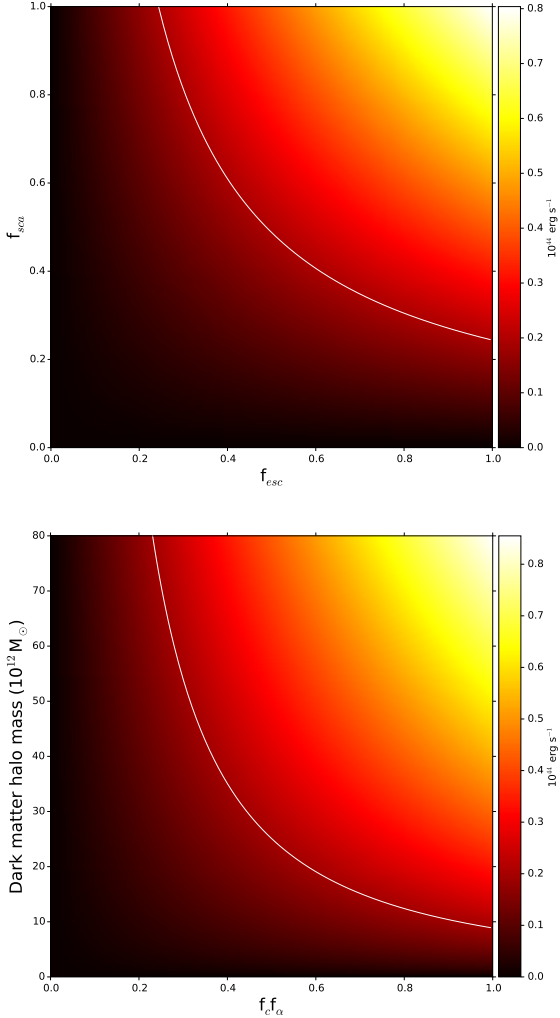

**Figure 5.** Upper: Colour maps showing how $L_{Ly\alpha}$ varies with $f_{esc}$ and $f_{sca}$ for the star formation model (upper) and $f_c f_\alpha$ and halo mass for the cold accretion model (lower), assuming $S_{850} = 0.6$mJy as obtained for the stacked LABs. The white line indicates the mean $L_{Ly\alpha}$ for all 34 LABs

Figure 5 (upper image) illustrates how $L_{Ly\alpha}$ (normalised to the $S_{850}$ values obtained for the full stack) varies with $f_{esc}$ and $f_{sca}$ for a star formation model. The mean $L_{Ly\alpha}$ of all 34 LABs, $\sim 2 \times 10^{43}$erg s$^{-1}$ (white line), can be obtained for a reasonable range of combinations of $f_{esc}$ and $f_{sca}$. The implications of the results in this section are discussed further in Section 5.

### 4.2 Cold mode accretion

We calculated theoretical $L_{Ly\alpha}$ for the cold accretion mode based on the toy model in Goerdt et al. (2010). The gravitational energy released to the cold gas as it streams from the virial radius towards the centre of the halo is expressed as:

$$\dot{E}_{grav} = f_c \dot{M}_c \hat{\phi} V_v^2 \qquad (7)$$

where $\dot{E}_{grav}$ is the gravitational power deposited in the cold gas at a given radius, r, per unit radial length, $f_c$ is the fraction of the total power produced that heats the cold stream (rather than heating the hot virialized gas, or increasing the velocity of the infalling gas), $\dot{M}_c$ is the accretion rate of the cold gas, $\hat{\phi}$ is the gravitational potential at r, and $V_v$ is the virial velocity.

Goerdt et al. (2010) considered an NFW potential well (Navarro et al. 1997), with a halo concentration parameter C$\approx$3 (C$= \frac{r_v}{r}$) for a halo of mass $10^{12} M_\odot$ at $z = 3$ (Bullock et al. 2001), resulting in $\hat{\phi} \approx 2.5$ as r $\rightarrow$ zero. For the purposes of our work we take the accretion rate to approximately equal the SFR, assuming that all the inflowing cold gas is converted to stars with no lag time. This gives:

$$\dot{E}_{grav} = f_c \times SFR \times 2.5 \times 236 \text{km}^{-2} M_{12}^{2/3} \qquad (8)$$

where virial velocity $\approx 236$kms$^{-1} M_{12}^{1/3}(1+z)_4^{1/2}$ (Goerdt et al. 2010), $M_{12} \equiv M_v/10^{12}M_\odot$ and $(1+z)_4 \equiv (1+z)/4$. A further parameter, $f_\alpha$ is added to represent the fraction of this energy emitted as observable Ly$\alpha$ radiation.

We used the Kennicutt & Evans (2012) SFR calibration for $L_{IR}$ (equation 2) , which assumes a Kroupa IMF, to produce SFRs for the range of $L_{IR}$ used in Section 4.1. In Fig. 4 (lower image) we plot theoretical lines for a $10^{12}M_\odot$, $10^{13}M_\odot$ and $10^{13.5}M_\odot$ dark matter halo (DMH) with $f_c f_\alpha = 0.34$. We used $f_c f_\alpha = 0.34$ as this was the value adopted by Goerdt et al. (2010) to obtain a reasonable fit to the observed LAB luminosity function (see section 5.3 for further discussion). As for the SF model, we also plot the individual LABs (pale grey), the stacked LABs (All 34 LABs), the area stacks (large LABs, medium LABs and small LABs) and the two individually detected LABs (LAB1 and 18). From the figure we see that LAB18 can easily be produced with a relatively low mass DMH, whereas LAB1, the full stack and the large LABs require a halo approaching $10^{13.5}M_\odot$. The small and medium LABs could possibly be produced with a slightly lower mass DMH.

Fig. 5 (lower image) illustrates how $L_{Ly\alpha}$ (normalised to the $S_{850}$ value obtained from the stack) varies with $f_c f_\alpha$ and halo mass for a cold accretion model. The mean $L_{Ly\alpha}$ of all 34 LABs, $\sim 2 \times 10^{43}$erg s$^{-1}$ (white line), can only be obtained with a massive DMH, $> 3 \times 10^{13}M_\odot$, for $f_c f_\alpha = 0.34$. The implications of the results in this section are discussed further in Section 5.

## 5 INTERPRETATION AND DISCUSSION

### 5.1 A summary of current detections in LABs

Our results show that on average the larger LABs ($> 29$ arcsec$^2$) are associated with submm emission, however there is no significant detection for the medium and small LABs. We summarise sources previously detected within the LABs in Table 3. X-ray detections in five of the LABs (Basu-Zych & Scharf 2004; Lehmer et al. 2009; Geach et al. 2009, BZ04, L09, G09) indicate the presence of an AGN which could easily provide the power needed to produce LABs (Cantalupo et al. 2012; Prescott et al. 2015). Weak radio detections were reported for only two of the LABs (Chapman et al. 2004; Geach et al. 2005, C04, G05), but these include LAB1, the largest and brightest LAB. IRAC detections in all four bands (Geach et al. 2007; Webb et al. 2009, G07, W09) have been obtained for 6 LABs, suggesting the presence of star forming, luminous galaxies hidden by dust. The two SCUBA-2 detections in our work also suggest the presence of dusty star forming galaxies. It is interesting to note that these luminous sources are only found in the larger LABs (> 21 arcsec²).

Table 3 also includes Lyman break galaxies (LBGs, Steidel et al. 2003, S03), the location of which within individual LABs was obtained from M04. These, together with the K-band NIR galaxies (Uchimoto et al. 2008; Uchimoto et al. 2012; Erb et al. 2014; Kubo et al. 2015; Kubo et al. 2016, U08, U12, E14, K15, K16) are common across all LABs, irrespective of size. Some of these *K*-band detections were associated with the LBGs, others were classified as Distant red galaxies (DRG). We also note two 1.1mm 3σ possible detections (Tamura et al. 2013, T13). We calculated the expected flux density for 1.1mm observations based on our $S_{850}$ = 0.6mJy stack and the modified black body model used in Section 4.1 as 0.3 mJy. This is consistent with the stacked value of $S_{1.1mm}$ <0.40 mJy (3σ) reported in (Tamura et al. 2013). More recently (Umehata et al. 2015, U15) have detected 1.1mm sources in LABs 12 and 14 (their deep ALMA observations were restricted to the central portion of the SSA22 protocluster).

The LBGs are not present in all LABs and therefore could not explain the observed Lyα emission in all cases. Furthermore the observed relationship between the $L_{IR}$ to $L_{UV}$ ratio and the UV spectral slope, β, (Meurer et al. 1999; Heinis et al. 2013; Álvarez-Márquez et al. 2015) suggests that the contribution from infrared sources will dominate that from LBGs.

The detections summarised above provide strong evidence that the larger LABs contain star forming galaxies, or AGN, that could produce, or at least significantly contribute to the observed LABs. The case for the smaller LABs is less convincing, as these do not contain such powerful counterparts. This split is consistent with our results from stacking by area and suggests that there may be two populations. It could be that the largest LABs are composed of multiple overlapping smaller LABs resulting from multiple sources. However, we should also note that the larger area of the large LABs group may increase the likelihood of chance alignment.

In addition, not all LABs have been observed at all wavelengths and deeper observations may reveal previously undetected sources in those that have. The upper limits for X-ray observations are generally low enough to make it unlikely that future observations would reveal an AGN, however LAB33 has an upper limit of 2 × 10⁴⁴ erg s⁻¹ allowing scope for a future detection. AGNs could also be obscured by dust (Geach et al. 2009), however the lack of IRAC 8μm detections in the smaller LABs makes this unlikely. The upper limits on IRAC and Submm observations do allow for future detections and ALMA observations would help to confirm the presence or otherwise of any dusty star forming galaxies in the smaller LABs and of multiple sources in the larger LABs.

### 5.2 Further exploration of powering mechanisms

We explored two options for powering the LABs in Section 4. We found that the full stack could be produced via star formation with

$f_{esc} f_{sca}$ = 0.2-0.3, or via cold accretion for a relatively massive halo (∼ 10¹³·⁵ $M_\odot$) with $f_c f_\alpha$ = 0.34. A value of $f_{esc} f_{sca}$ = 0.2 (0.3) implies that 20 per cent (30 per cent) of the Lyα emitted within a central galaxy is able to escape from the galaxy and is then scattered by the CGM into our line of sight. This value is higher than the figure of 0.1 typically found in the literature (Hayes et al. 2011c; Dijkstra & Jeeson-Daniel 2013), but these figures generally apply to LAEs. For LAEs any scattering takes place close to the galaxy, however, in the case of LABs, scattering is occurring over an extended region and a higher value of $f_{sca}$ is therefore plausible. There is also some variation in escape fraction values in the literature. Wardlow et al. (2014) found a significantly higher range of values, with lower limits of 0.1-0.3 depending on the SED used to fit their LAEs. Geach et al. 2009 found that a value as low as 0.006 was sufficient to power the LABs containing AGN. However, if their figure 4 were recreated using our SCUBA-2 results, this would result in a higher escape fraction, which is more in line with our results. In order for Lyα to escape from the central galaxy there must be a relatively low covering fraction (Steidel et al. 2011; Trainor et al. 2015). However, sufficient cold gas is also required in the CGM in order for scattering to take place. This suggests the need for an irregular, patchy CGM (Steidel et al. 2010), in line with predictions from recent simulations (van de Voort et al. 2011; van de Voort & Schaye 2012; Rosdahl & Blaizot 2012). The dark matter halo function obtained from simulations (Springel et al. 2005; Lukić et al. 2007; Martinovic 2015), suggests that a ∼10¹³·⁵ $M_\odot$ DMH is rare at $z$ ∼3. Whilst such a massive DMH might be possible for some of the LABs (e.g LAB1) it is not possible for all LABs to exist in such massive DMHs. Thus our cold accretion model appears to require more extreme conditions than the star formation (SF) model and we suggest that, on average, star formation is more likely to be the dominant process. Additional contributions from cold accretion and AGN are also likely for some individual LABs.

LAB18 is relatively easy to produce via either process, but LAB1 requires more extreme conditions; either a high mass halo or a higher value of $f_{esc} f_{sca}$ than proposed by (Hayes et al. 2011c; Dijkstra & Jeeson-Daniel 2013). In fact, as can be seen from Table 3, LAB1 contains multiple galaxies and a radio source. It is therefore likely that this LAB is the result of more than one process and a higher value of $f_{esc} f_{sca}$ may not be required to explain the observed emission.

The results for the large LABs stack are similar to those for the full stack, though not quite as extreme. The value of $f_{esc} f_{sca}$ = 0.2 is more achievable than the extremely massive dark matter halo that is required under the cold accretion model. This makes SF more likely to be the dominant process for the large LABs, although contributions from other processes may also be required. The results for the medium and small LABs require a lower $f_{esc} f_{sca}$, consistent with the literature and so these smaller LABs could be produced from SF without contributions from other sources. However, they also require a lower mass DMH (∼10¹³ $M_\odot$) than other stacks and therefore this scenario is also possible.

Our results indicate a significantly lower mean $S_{850}$ than G05. This implies a smaller chance of finding submm sources in all LABs and lower SFRs (∼50 $M_\odot$ rather than ∼10³ $M_\odot$). However, our analysis still favours central star formation over cold accretion as the primary fuelling method. G05 considered the superwind model rather than the two models discussed in this paper. We have not repeated this aspect of their work as alternative models involving scattering are now considered more likely (Hayes et al. 2011b; Geach et al. 2014). G05 found a trend between the $L_{Ly\alpha}$ of the haloes and submm flux, but we find only a marginal correla-

**Table 3.** Summary of sources detected in SSA22 LABs based on published literature.

| ID | X-ray $(L_{2-32keV})^a$ $10^{44}$ erg s$^{-1}$ | IRAC 3.6$^b$ μJy | IRAC 4.5$^b$ μJy | IRAC 5.8$^b$ μJy | IRAC 8.0$^b$ μJy | SCUBA-2$^c$ mJy | NIR$^d$ K$_{AB}$ | Notes$^e$ |
|---|---|---|---|---|---|---|---|---|
| LAB1 | <0.24 | 7.3±0.1 | 9.2±0.2 | 11.5±0.8 | 14.2±1.0 | 4.6±1.1 | 22.97±0.11 (DRG*) | Radio 21cm, MIPS |
| | | 8.4±0.1 | 11.1±0.2 | 14.7±0.8 | 15.9±1.0 | | 23.34±0.11 (C11*) | LBGs (C11,C15) |
| | | | | | | | 22.02±0.10 | |
| | | | | | | | 23.97±0.22 | |
| LAB2 | 0.81±0.03 | 5.6±0.2 | 6.7±0.2 | 8.3±0.9 | 6.5±1.4 | <3.9 | 22.91±0.11 (DRG, | LBG (M14) |
| | | 5.8±0.2 | 7.8±0.2 | 10.5±0.9 | 15.4±1.4 | | vicinity M14*) | |
| LAB3 | 2.13±0.02 | | | | | <3.9 | 22.54±00.12 (DRG) | |
| LAB4 | <0.56 | | | | | <3.9 | | Radio 21cm |
| LAB5 | <0.44 | 7.7±0.1 | 9.4±0.2 | 10.1±1.0 | 12.4±1.3 | <3.9 | <24.3 | |
| LAB6 | | 5.2±0.3 | 5.4±0.8 | <9.0 | <25.0 | <3.9 | | |
| LAB7 | <0.22 | 2.8±0.4 | 2.8±0.3 | <7.0 | <10.0 | <3.9 | 23.49±0.12 (DRG, C6*) | LBGs (C6,M4) |
| | | | | | | | 23.65±0.13 (M4*) | |
| | | | | | | | 23.97±0.18 (C6*) | |
| LAB8 | <0.20 | 2.6±0.1 | 1.0±0.2 | <4.0 | <5.0 | <3.9 | 24.35 ±0.28 (C15*) | LBG (C15) |
| LAB9 | <0.37 | 2.3±0.2 | 2.2±0.2 | <5.0 | <10.5 | <3.9 | <24.3 | |
| LAB10 | | | | | | <3.9 | | |
| LAB11 | <0.28 | 3.0±0.2 | 2.6±0.2 | <5.0 | <6.0 | <3.9 | 23.79±0.17 (C47*) | LBG (C47) |
| | | | | | | | 24.49±0.23 (C47*) | |
| | | | | | | | 24.94±0.24 (C47*) | |
| LAB12 | 0.91±0.03 | 2.6±0.2 | 3.0±0.2 | 4.8±0.9 | <6.0 | <3.9 | 22.11±0.10 | LBG (M28) |
| | | | | | | | 23.67±0.21 (M28*) | |
| LAB13 | <1.57 | | | | | <3.9 | | $S_{1.1}$ = 0.7±0.1 mJy U15 |
| LAB14 | 1.82±0.02 | 7.4±0.1 | 10.6±0.2 | 15.0±0.8 | 19.6±1.1 | <3.9 | 22.45±0.11 (DRG) | $S_{1.1}$ = 1.8±0.1 mJy U15 (3σ, T13), MIPS |
| LAB15 | <0.87 | 1.7±0.3 | 1.8±0.3 | <12.0 | <8.0 | <3.9 | | |
| LAB16 | <0.36 | 5.8±0.1 | 7.3±0.2 | 9.6±1.1 | 15.2±1.5 | <3.9 | 22.99±0.11 | MIPS |
| | | | | | | | 23.51±0.19 | |
| LAB17 | | | | | | | | |
| LAB18 | 1.59±0.03 | 7.3±0.2 | 8.7±0.3 | 15.7±1.5 | 19.2±1.6 | 5.2±1.1 | | MIPS, 1.1mm (3σ, T13) |
| | | 4.9±0.2 | 7.8±0.3 | 8.6±1.5 | 17.6±1.6 | <3.9 | | |
| LAB19 | <0.36 | 2.8±0.1 | 2.4±0.2 | <4.0 | <5.5 | <3.9 | <24.3 | |
| LAB20 | <0.22 | 1.3±0.2 | <1.0 | <4.0 | <7.5 | <3.9 | 23.3±0.11 (C12*) | LBG (C12) |
| LAB21 | <2.84 | | | | | <3.9 | | |
| LAB22 | <0.56 | <3.0 | <2.0 | <6.0 | <8.0 | <3.9 | | |
| LAB23 | | <1.5 | <4.0 | <13.0 | <27 | <3.9 | | |
| LAB24 | <0.39 | <0.8 | <1.0 | <5.0 | <7.0 | <3.9 | 23.36±0.13 (DRG) | |
| LAB25 | <0.23 | 1.2±0.1 | 2.4 | <4.0 | <8.0 | <3.9 | <24.3 | |
| LAB26 | <0.23 | <1.0 | <1.0 | <5.0 | <10.0 | <3.9 | <24.3 | |
| LAB27 | <0.97 | | | | | <3.9 | <24.3 | |
| LAB28 | <0.99 | | | | | <3.9 | | |
| LAB29 | | 3.6±0.3 | <4.0 | <8.0 | <24.0 | <3.9 | | |
| LAB30 | <0.27 | | | | | <3.9 | 22.96±0.11 (D3*) | LBG (D3) |
| | | | | | | | 23.51±0.19 | |
| LAB31 | <0.23 | 1.1±0.3 | <1.5 | <6.0 | <8.0 | <3.9 | 23.20±0.11 (C16*) | LBG (C16) |
| LAB32 | <0.47 | | | | | <3.9 | | |
| LAB33 | <2.08 | 1.8±0.2 | 2.1±0.3 | <10.0 | <10.0 | <3.9 | | |
| LAB34 | | 1.8±0.2 | <2.0 | <6.5 | <13.5 | <3.9 | | |
| LAB35 | <0.27 | 1.4±0.2 | 1.4±0.2 | <4.0 | | <3.9 | <24.3 | |

a) Luminosities taken from G09, originally reported in L09, first detection of LAB2 in BZ04.

b) IRAC flux densities taken from W09, original detections for LAB1,2 in G07, further information on LAB1 in G14. Some LABs contain more than one source.

c) This work.

d) AB magnitudes from U12 (many originally detected in U08). U12 used a 6″ radius and a combination of photometric redshifts and Lyman-alpha spectroscopic redshifts ($z_{spec}$) of probable K band counterparts to identify LAB sources (* indicates $z_{spec}$ was provided in U12). In some cases there are multiple possible K band counterparts for a single LBG. Where no LAB counterpart was detected by U12 an upper limit is quoted. DRG stands for Distant Red Galaxy. U12 covered only 20 of the LABs, there is currently no information available for the other 15. NIR $z_{spec}$ were later obtained for the counterparts of LABs 1,12,12 30 and 35 (E14,U15,U16) and for these LABS these values are used to determine membership of the protocluster.

e) LBGs are listed together with their ID numbers from S03, their positions within the LABs are taken from M04. MIPS detections are from G07 and Radio 21cm detections from C04 and G05.

tion to luminosity and a much higher correlation to the size of the LABs.

### 5.3 Underlying assumptions

We now consider some of the assumptions made in applying these models. In our SF model we assume a dust temperature of 30K which compares to an average measured value for submm galaxies of 30-40K (Casey et al. 2014). We tested the sensitivity of our work to dust temperature by increasing the temperature to 40K and found that this increased the $L_{IR}$ by $\sim 4 \times 10^{45}$ ergs$^{-1}$. Such an increase allows LAB1 the full stack and large LABs to be created with $f_{esc}f_{sca} = 0.05 - 0.1$, in line with values in the literature. We also varied the value of $\beta$ in our model (usually found to be in the range 1-2 in starburst galaxies (Hildebrand 1983; Chapin et al. 2011; Casey et al. 2014)). For a value of $\beta = 1$ (and 30K) $L_{IR}$ is decreased by $\sim 1 \times 10^{45}$ ergs$^{-1}$, requiring $f_{esc}f_{sca} > 0.3$ to generate LAB1 the full stack and large LABs. It is therefore possible that with a high dust emissivity and temperature the SF model could account for all the Lyman-$\alpha$ emission observed in the LABs. However, a lower value of $\beta$ together with a low temperature make the SF model less likely as the sole source of emission.

The value of $C$ used to compute the cold accretion Ly$\alpha$ values in section 4.2 is based on the Bullock et al. (2001) formula, $C \approx 3M_{12}^{-0.13}(1+z)_4^{-1}$, which results in $C \approx 3$ for a halo mass of $10^{12}M_\odot$ at $z = 3$. However, a more massive halo may be required to produce the LABs by cold accretion. Assuming a halo mass of $10^{13}M_\odot$ results in a revised value of $\hat{\phi} \approx 2.2$. Re-plotting Fig. 4 using this revised value of $\hat{\phi}$ had no significant impact on our conclusions. We also assume that the accretion rate is approximately equal to the SFR. If only a proportion of the accreted gas produces stars then we are underestimating the accretion rate in our model and therefore underestimating the production of Ly$\alpha$ emission due to cold accretion. This could allow for the production of LAB1 and the stack with a lower halo mass.

The value of $f_c f_\alpha$ (the fraction of gravitational power released that heats the cold streams $\times$ the fraction of energy emitted as observable Ly$\alpha$ emission) used in the cold accretion model is uncertain. We use 0.34 as this was the value adopted by (Goerdt et al. 2010) to obtain a reasonable fit to the observed LAB luminosity function. However, in the same paper, a figure of $f_\alpha = 0.85$ is used for their hydrodynamic simulations. This would require $f_c = 0.4$ if applied to the toy model, but Goerdt et al. (2010) suggest elsewhere that $f_c \sim 1$. There is disagreement in the literature as to whether the infall velocity of the cold streams is close to free fall, giving a value of $f_c \sim 0$ (Rosdahl & Blaizot 2012; Faucher-Giguère et al. 2010), or whether the velocity is largely constant for all radii, giving $f_c \sim 1$ (Goerdt et al. 2010; Goerdt & Ceverino 2015). Rosdahl & Blaizot (2012) found $f_c$ between 0 and 0.3 in AMR zoom simulations, with a higher value in messy streams in low mass haloes. By contrast, the Goerdt & Ceverino (2015) simulations indicated a constant velocity in low mass haloes and only a slight increase in velocity for higher mass haloes ($> 5 \times 10^{12}M_\odot$). If we were to use a maximum value of $f_c f_\alpha = 1 \times 0.85 = 0.85$ in Fig. 4 LAB1 and the full stack could be produced with a halo mass of $10^{13}M_\odot$, still massive, but less extreme than required with $f_c f_\alpha = 0.34$. Given the large uncertainties in the submm flux LAB1 could potentially be fuelled in a $10^{13}M_\odot$ halo with $f_c f_\alpha$ as low as 0.55, but this would not produce the flux seen in the stack or large LABs.

The non detection of a submm source does not necessarily rule out the presence of a luminous galaxy. A luminous source could be obscured by dust in cold clumpy gas, which would not be observable from some orientations.

Finally, we note that in both scenarios cold gas is required outside the galaxies in order to scatter the Ly$\alpha$ into our line of sight. Therefore, even if the Ly$\alpha$ emission is fuelled primarily by star formation processes, the observed LABs are still evidence of the presence of cold gas in the CGM. In addition the high value of $f_{esc}f_{sca}$ implied suggests that a clumpy configuration is more likely than narrow streams in the CGM and requires inhomogeneous HII regions within the galaxy powering the LAB (Haiman & Spaans 1999).

## 6   CONCLUSIONS

We used SCUBA-2 CLS data to search for SMGs in the SSA22 LABs, investigating both individual detections and the mean stack. We then compared our results to two potential mechanisms for fuelling LABs, central star formation and cold accretion. Our findings are as follows:

● Two of the LABs had SCUBA-2 detections at $> 3.5\sigma$, LAB1 (4.6±1.1mJy) and LAB18 (5.2±1.1mJy).

● Our mean stacking of all 34 LABs resulted in an average flux density of 0.6±0.2mJy at 3.1$\sigma$. This implies that, on average, each LAB contains a dusty, star forming galaxy and that star formation processes are at least partly responsible for fuelling the LABs.

● Our stacking of the LABs based on their size suggests that the larger LABs are marginally more likely to contain a submm source ($S_{850} > 1$mJy) than the smaller LABs. A review of the literature suggests that luminous sources can be found in most of the larger LABs (LABs 1–18), but are missing from the smaller LABs. Whilst future observations may change this picture, it is possible that there are two populations of LABs, large LABs created by luminous sources (galaxies or AGN) and smaller LABs containing less luminous galaxies, or fuelled solely by cold accretion.

● Our investigation of two possible fuelling processes suggests that central star formation is more consistent with being the dominant source of the Ly$\alpha$ emission than cold accretion. However, given the uncertainty in $f_c f_\alpha$ and $f_{esc}f_{sca}$, neither process can be ruled out and it is likely that both processes are involved to some extent in most LABs.

Deeper data from the Atacama Large Millimeter/submillimeter Array (ALMA) are required to detect individual SMGs in the LABs and place further constraints on the fuelling processes.

## ACKNOWLEDGEMENTS

We thank Ian Smail for useful discussions and comments. N.K.H. and M.J.M are supported by the Science and Technology Facilities Council [N.K.H. grant number ST/K502029/1]. J.E.G. is supported by a Royal Society University Research Fellowship. Y.M. acknowledges support from Japan Society for the Promotion of Science KAKENHI Grant Number 20647268. The James Clerk Maxwell Telescope has historically been operated by the Joint Astronomy Centre on behalf of the Science and Technology Facilities Council of the United Kingdom, the National Research Council of Canada and the Netherlands Organisation for Scientific Research. Additional funds for the construction of SCUBA-2 were provided by the Canada Foundation for Innovation. This research made use of the NASA/ IPAC Infrared Science Archive, which is operated by the Jet Propulsion Laboratory, California Institute of Technology, under contract with the National Aeronautics and Space Administration and of the NASA's Astrophysics Data System. This research also made use of the following software packages: The STARLINK software (Currie et al. 2014), supported by the East Asian Observatory; the IPYTHON package (Pérez & Granger 2007); MATPLOTLIB (Hunter 2007); SCIPY (Jones & et al. 2001); MONTAGE (NASA/IPAC); the KAPETYN Package (Terlouw & Vogelaar 2012); ASTROPY (Astropy Collaboration et al. 2013); COSMOLOPY. The authors wish to recognize and acknowledge the very significant cultural role and reverence that the summit of Mauna Kea has always had within the indigenous Hawaiian community. We are most fortunate to have the opportunity to conduct observations from this mountain.

This paper has been typeset from a TeX/LaTeX file prepared by the author.